\documentclass[aps,prl,preprint,showpacs,byrevtex]{revtex4}

\usepackage{times,mathptm}
\usepackage[dvips]{graphicx,color}

\begin{document}

\title{Driving Force of Phase Transition in Indium Nanowires on Si(111)}
\author{Hyun-Jung Kim and Jun-Hyung Cho$^{*}$}
\affiliation{Department of Physics and Research Institute for Natural Sciences, Hanyang University,
17 Haengdang-Dong, Seongdong-Ku, Seoul 133-791, Korea}

\date{\today}

\begin{abstract}
The precise driving force of the phase transition in indium nanowires on Si(111) has been controversial whether it is driven by a Peierls instability or by a simple energy lowering due to a periodic lattice distortion. The present van der Waals (vdW) corrected hybrid density functional calculation predicts that the low-temperature 8${\times}$2 structure whose building blocks are indium hexagons is energetically favored over the room-temperature 4${\times}$1 structure. We show that the correction of self-interaction error and the inclusion of vdW interactions play crucial roles in describing the covalent bonding, band-gap opening, and energetics of hexagon structures. The results manifest that the formation of hexagons occurs by a simple energy lowering due to the lattice distortion, not by a charge density wave formation arising from Fermi surface nesting.
\end{abstract}


\maketitle


One-dimensional (1D) electronic systems have attracted much attention because of the richness of exotic physical phenomena such as charge density wave formation due to the  Peierls instability~\cite{peierls}, non-Fermi liquid behavior~\cite{stewart}, or Jahn-Teller distortion~\cite{jahn}. A prototypical example of quasi-1D systems is self-organized indium nanowires on the Si(111) surface~\cite{snijders,bunk,yeom1}. Each nanowire is composed of two zigzag chains of In atoms, and the nanowires are separated by a zigzag chain of Si atoms (see Fig. 1)~\cite{bunk}. Below ${\sim}$120 K, this quasi-1D system undergoes a reversible phase transition initially from a 4${\times}$1 structure to a 4${\times}$2 structure, then to an 8${\times}$2 structure~\cite{yeom1,kump}, showing a period doubling both parallel and perpendicular to the In wires. This (4${\times}$1)${\leftrightarrow}$(8${\times}$2) phase transition is accompanied by a metal-insulator transition~\cite{yeom1,ahn,yeom2}. These intriguing results have stimulated many experimental~\cite{park,ahn,gon,chandola} and theoretical studies~\cite{cho1,cho2,tsay,lopez,gon1,gon2,cho3,rii,steko}. However, the precise driving force of the phase transition has been elusive for a long time. It has been suggested that the phase transition is driven by a Peierls instability~\cite{yeom1,ahn,yeom2,park} or by a simple energy lowering due to a periodic lattice distortion~\cite{cho1,cho2,tsay,lopez,gon1,gon2,rii,steko}. The former mechanism involves the strong coupling between lattice vibrations and electrons near the Fermi level caused by Fermi surface nesting. Consequently, the charge density wave formation together with the lattice distortion occurs because of a larger electronic energy gain compared to an elastic energy cost. On the other hand, the latter mechanism involves either the trimer formation~\cite{cho1,cho2,tsay,lopez} in In chains with an elastic energy gain arising from the lattice distortion or the hexagon formation~\cite{cho3,gon1,gon2,rii,steko} with an elastic energy gain from the lattice distortion as well as an electronic energy gain from the band-gap opening.

\begin{figure}[ht]
\centering{ \includegraphics[width=7.0cm]{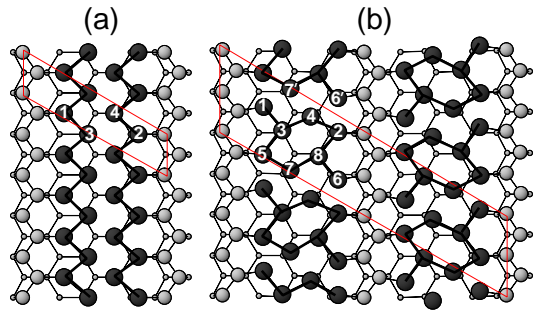} }
\caption{Top view of the optimized (a) 4${\times}$1 and (b) 8${\times}$2 structures of the In/Si(111) surface system. The dark and gray circles represent In and Si atoms, respectively.
For distinction, Si atoms in the subsurface are drawn with small circles. Each unit cell is indicated by the solid line.}
\end{figure}

Despite the above-mentioned controversial issue on the origin of the phase transition in the In/Si(111) system, the so-called hexagon model describes well several observed features of the low-temperature phase such as an insulating character~\cite{yeom1,ahn,yeom2}, scanning tunneling microscopy images~\cite{gon}, and anisotropic optical interband transitions~\cite{chandola}. Initially, Gonz$\acute{\rm a}$lez, Ortega, and Flores proposed that a shear distortion,
whereby neighboring In chains are displaced in opposite
directions, allows for the formation of hexagon in the 4${\times}$2 unit cell~\cite{gon1,gon2}. Since this shear phonon mode~\cite{gon} is different from a phonon mode with the observed~\cite{yeom1} Fermi surface nesting vector 2$k_{F}$ = ${\pi}$/$a_{\rm x}$ ($a_{\rm x}$ is the lattice constant along the In chains), the Peierls mechanism is unlikely to be the driving force of the phase transition in the In/Si(111) system. Moreover, the stabilization of the 8${\times}$2 structure by doubling the unit cell perpendicular to the In wires is irrelevant with an electron-phonon coupling due to the observed Fermi surface nesting along the direction parallel to the In wires. The density-functional theory (DFT) calculations of Gonz$\acute{\rm a}$lez, Ortega, and Flores showed that the 4${\times}$2 or 8${\times}$2 hexagon structure is energetically favored over the 4${\times}$1 structure~\cite{gon1,flores}, but subsequent more accurate DFT calculations~\cite{steko,cho3} within the local density~\cite{ca} as well as generalized gradient approximation~\cite{pw} (LDA/GGA) did not support the energetic preference of the hexagon structures (see Table I). According to an LDA calculation with keeping the In 4$d$ electrons frozen, the 8${\times}$2 hexagon structure was predicted to be energetically favored over the 4${\times}$1 structure~\cite{steko}. However, this result is an artifact of the relatively less accurate scheme. Because of the proper prediction for the energy stability between the 4${\times}$1 and 8${\times}$2 structures, the LDA scheme with a frozen core of In 4$d$ electrons has been employed to calculate the entropy difference~\cite{wippermann} or the energy barrier between the two structures~\cite{wall}. We note, however, that the LDA and GGA calculations with the treatment of the In 4$d$ states as valence electrons predicted that the 4${\times}$2 and 8${\times}$2 hexagon structures are less stable than the 4${\times}$1 structure~\cite{steko}.

In this Letter, we present a new theoretical study which extends the previous work by considering a hybrid exchange-correlation functional~\cite{hse} and by taking van der Waals (vdW)~\cite{vdw} interactions into account. We will show that the correction of self-interaction error (SIE) cures over-delocalization of surface-state electrons inherent in the DFT and therefore describes adequately the covalent bonding, band-gap opening, and energetics of hexagon structures. Furthermore, we find that the vdW interactions between In atoms play an important role in further stabilizing the 4${\times}$2 and 8${\times}$2 hexagon structures. Since the formation of hexagons and the more stabilization of the 8${\times}$2 structure are not associated with an electron-phonon coupling due to Fermi surface nesting, we can say that the phase transition in the In/Si(111) system is driven by a simple energy lowering due to the hexagon formation rather than by a Peierls-like mechanism.

The present vdW corrected hybrid DFT calculations were performed using the FHI-aims~\cite{aims} code for an accurate, all-electron description based on numeric atom-centered orbitals, with ``tight" computational settings. For the exchange-correlation energy, we employed the hybrid functional of HSE~\cite{hse} as well as the GGA functional of PBE~\cite{per}. The ${\bf k}$-space integrations in various unit-cell calculations were done equivalently with 64 ${\bf k}$ points in the surface Brillouin zone of the 4${\times}$1 unit cell. The Si(111) substrate (with the Si lattice constant $a_0$ = 5.482 {\AA}) was modeled by a 6-layer slab (not including the Si surface chain bonded to the In chains) with ${\sim}$15 {\AA} of vacuum in between the slabs. Each Si atom in the bottom layer was passivated by one H atom. All atoms except the bottom layer were allowed to relax along the calculated forces until all the residual force components were less than 0.02 eV/{\AA}.

\begin{table*}[ht]
\caption{Calculated total energies (in meV per 4${\times}$1 unit cell) of the 4${\times}$2 and 8${\times}$2 structures relative to the 4${\times}$1 structure, together with the band gaps (in eV). For comparison, the previous LDA~\cite{ca} and GGA~\cite{pw} results are also given. ``Valence $d$ (core $d$)" represents the treatment of the In 4d states as valence (core) electrons.}
\begin{ruledtabular}
\begin{tabular}{lcccc}
& \multicolumn{2}{c}{4${\times}$2 } & \multicolumn{2}{c}{8${\times}$2 }  \\
\cline{2-3}
\cline{4-5}
                              &   ${\Delta}E$     &  $E_{\rm g}$   &       ${\Delta}E$     &  $E_{\rm g}$   \\ \hline
 PBE                        &       33     &  no   &       26     &  0.08  \\
 PBE+vdW                    &       22     &  0.05 &       13     &  0.08  \\
 HSE                        &        3     &  0.10 &      $-$15     &  0.19  \\
 HSE+vdW                    &      $-$23     &  0.21 &      $-$40     & 0.21  \\
 LDA (Ref. ~\cite{gon1,gon2})    &   $-$80  &  &  $-$100 &   \\
 GGA$-$valence $d$ (Ref. ~\cite{steko})             &       48     &      &       27     &   0.05     \\
 GGA$-$core $d$ (Ref. ~\cite{steko})             &       36     &       &       25     &        \\
 LDA$-$valence $d$ (Ref. ~\cite{steko})             &       15     &       &       2     &        \\
 LDA$-$core $d$ (Ref. ~\cite{steko})             &       5     &       &       $-$12     &        \\
\end{tabular}
\end{ruledtabular}
\end{table*}

We begin to optimize the 4${\times}$1, 4${\times}$2, and 8${\times}$2 structures using the PBE functional. The optimized 4${\times}$1 and 8${\times}$2 structures are displayed in Fig. 1(a) and 1(b), respectively. We find that the 4${\times}$2 and 8${\times}$2 structures show the formation of hexagons. Unlike previous pseudopotential calculations~\cite{cho1,cho2,steko}, the present all-electron calculations were not able to find the stabilization of trimers in the 4${\times}$2 and 8${\times}$2 structures, which were converged to the 4${\times}$1 structure. The calculated total energies (in meV per 4${\times}$1 unit cell) of the 4${\times}$2 and 8${\times}$2 structures relative to the 4${\times}$1 structure are given in Table I. We find that the 4${\times}$2 and 8${\times}$2 structures are less stable than the 4${\times}$1 structure with ${\Delta}E_{\rm 4{\times}2-4{\times}1}$ = 33 meV and ${\Delta}E_{\rm 8{\times}2-4{\times}1}$ = 26 meV, respectively, consistent with those (48 and 27 meV in Table I) obtained by a previous GGA calculation~\cite{steko} with the Perdew-Wang exchange-correlation functional~\cite{pw}. The calculated interatomic distances of In atoms are given in Table II. In the 4${\times}$1 structure, the In-In distance $d_{{\rm In}_{1}-{\rm In}_{3}}$ ($d_{{\rm In}_{2}-{\rm In}_{4}}$) within an In chain is 3.045 (3.047) {\AA}, while $d_{{\rm In}_{3}-{\rm In}_{4}}$ between the two In chains is 3.115 {\AA}. However, in the 4${\times}$2 (8${\times}$2) structure, the In-In distances between the two In chains are shortened as $d_{{\rm In}_{3}-{\rm In}_{4}}$ = 3.054 (3.030) and $d_{{\rm In}_{7}-{\rm In}_{8}}$ = 3.027 (3.011) {\AA}, leading to the formation of hexagon [see Fig. 1(b)]. We note that each In-In distance in the 8${\times}$2 structure slightly changes compared to the corresponding one in the 4${\times}$2 structure because of the formation of hexagons in two opposite orientations (see Table II).

\begin{table*}[ht]
\caption{Calculated interatomic distances (in {\AA}) of In atoms in the 4${\times}$1, 4${\times}$2, and 8${\times}$2 structures using PBE. The results obtained using PBE+vdW are also given in parentheses. The labeling of In atoms are shown in Fig. 1.}
\begin{ruledtabular}
\begin{tabular}{llcccccc}
            && 4${\times}$1 && 4${\times}$2 && 8${\times}$2 \\ \hline
   In$_1$$-$In$_3$   &&  3.045 (3.040)&&   2.949 (2.904)  && 2.941 (2.912)    \\
   In$_2$$-$In$_4$   &&  3.047 (3.053)&&  2.999 (3.006)   && 2.996 (2.994)      \\
   In$_3$$-$In$_4$   &&  3.115 (3.092)&&  3.054 (3.041)   && 3.030 (3.017)      \\
   In$_3$$-$In$_5$   &&                       &&  3.018 (2.993)   && 2.999 (2.975)      \\
   In$_2$$-$In$_8$   &&                       &&  3.024 (3.005)   && 2.995 (2.983)      \\
   In$_5$$-$In$_7$   &&                       &&  3.010 (3.021)   && 3.028 (3.026)      \\
   In$_8$$-$In$_6$   &&                       &&  2.952 (2.913)   && 2.952 (2.928)      \\
   In$_7$$-$In$_8$   &&                       &&  3.027 (3.029)   && 3.011 (3.011)      \\

   In$_1$$-$In$_{7'}$   &&                    &&  3.171 (3.201)   && 3.128 (3.153)      \\
   In$_4$$-$In$_{7'}$   &&                    &&  3.294 (3.306)   && 3.347 (3.348)      \\
   In$_4$$-$In$_{6'}$   &&                    &&  3.130 (3.178)   && 3.119 (3.160)      \\
\end{tabular}
\end{ruledtabular}
\end{table*}

To examine the influence of vdW interactions on the geometry and energetics, we use the PBE+vdW scheme developed by Tkatchenko and Scheffler~\cite{vdw}, where the vdW coefficients and radii are determined using the self-consistent electron density~\cite{guo}. As shown in Table II, several In-In distances obtained using PBE and PBE+vdW exhibit some differences by less than 0.05 {\AA}. The calculated PBE+vdW total energies of the 4${\times}$2 and 8${\times}$2 structures relative to the 4${\times}$1 structure are also listed in Table I. We find that the 4${\times}$2 and 8${\times}$2 structures are still less stable than the 4${\times}$1 structure with ${\Delta}E_{\rm 4{\times}2-4{\times}1}$ = 22 meV and ${\Delta}E_{\rm 8{\times}2-4{\times}1}$ = 13 meV, respectively. Thus, the inclusion of vdW interactions within the PBE+vdW scheme does not reverse the stability of the 4${\times}$2 (or 8${\times}$2) and 4${\times}$1 structures.

The calculated surface band structures of the 4${\times}$2 structure obtained using PBE and PBE+vdW are displayed in Fig. 2(a) and 2(b), respectively. It is seen that PBE gives a metallic feature while PBE+vdW gives an insulating feature with a band gap opening ($E_{\rm g}$) of 0.05 eV. The PBE and PBE+vdW calculations for the 8${\times}$2 structure give almost the same value of $E_{\rm g}$ = 0.08 eV [see Fig. 1(a) and 1(b) of the Supplemental Material (SI)]. For the 4${\times}$1 structure, both PBE and PBE+vdW predict well the observed metallic feature~\cite{ahn}, where three surface bands cross the Fermi level [see Fig. 2(a) and 2(b) of the SI]. Thus, we can say that PBE cannot predict the observed~\cite{yeom1,ahn,yeom2} insulating feature for the low-temperature phase, consistent with a previous pseudopotential calculation~\cite{cho3}.

\begin{figure}[ht]
\centering{ \includegraphics[width=7.0cm]{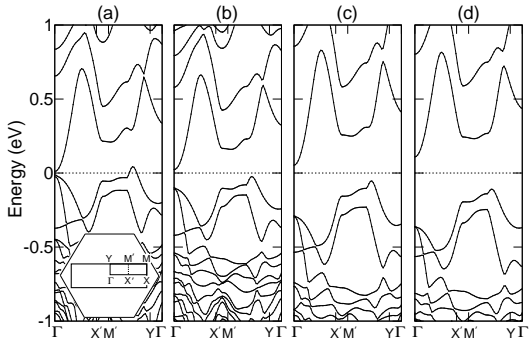} }
\caption{Surface band structure of the 4${\times}$2 structure obtained using (a) PBE, (b) PBE+vdW, (c) HSE, and (d) HSE+vdW. The inset in (a) shows the surface Brillouin zone for the 4${\times}$1 and 4${\times}$2 unit cells within that for the 1${\times}$1 unit cell. The energy zero represents the Fermi level.}
\end{figure}

The local and semi-local DFT have challenged for a reasonable description of the energetics of different structures involved in the phase transition or the kinetics of chemical reactions because of their intrinsic SIE~\cite{sie1,sie2}. Especially, the GGA tends to stabilize artificially delocalized states due to the SIE, since delocalization reduces the self-repulsion. It is thus likely that the present PBE functional would give a lower energy for the metallic 4${\times}$1 structure, compared to the 4${\times}$2 and 8${\times}$2 structures. In order to correct the SIE, we use the hybrid HSE functional~\cite{hse} to calculate the total energies of the 4${\times}$1, 4${\times}$2, and 8${\times}$2 structures with the PBE geometries~\cite{relax}. We find that the correction of SIE enhances the stability of the 4${\times}$2 and 8${\times}$2 structures relative to the 4${\times}$1 structure, giving rise to ${\Delta}E_{\rm 4{\times}2-4{\times}1}$ = 3 meV and ${\Delta}E_{\rm 8{\times}2-4{\times}1}$ = $-$15 meV, respectively. This enhanced HSE stability of the 4${\times}$2 (8${\times}$2) structure is caused by the electronic energy gain arising from an increased band gap of $E_{\rm g}$ = 0.10 (0.19) eV, as shown in Fig. 2(c) [Fig. 1(c) of the SI]. Thus, HSE predicts well the observed~\cite{yeom1,ahn,yeom2} insulating feature for the 4${\times}$2 and 8${\times}$2 structures.

To see the effects of the SIE on the charge density distribution, we plot the charge density difference defined as
\begin{equation}
{\Delta}{\rho} = {\rho}_{\rm HSE} - {\rho}_{\rm PBE},
\end{equation}
where ${\rho}_{\rm HSE}$ (${\rho}_{\rm PBE}$) is the charge density obtained using the HSE (PBE) functional. The results for the 4${\times}$1 and 4${\times}$2 structures are displayed in Fig. 3(a) and 3(b), respectively. We find a conspicuous difference between the 4${\times}$1 and 4${\times}$2 structures for ${\Delta}{\rho}$. It is seen that the insulating 4${\times}$2 structure has a larger ${\Delta}{\rho}$ between In atoms compared with the metallic 4${\times}$1 structure, indicating that the relatively localized surface states in the former are more affected by the SIE than the delocalized surface states in the latter. This fact also reflects that in the 4${\times}$2 structure, the correction of the SIE by HSE recovers the charge localization in the covalently bonding between In atoms. As a consequence, we obtain an increase of band gap with $E_{\rm g}$ = 0.10 eV, leading to a decrease of ${\Delta}E_{\rm 4{\times}2-4{\times}1}$ = 3 meV compared to the PBE result (${\Delta}E_{\rm 4{\times}2-4{\times}1}$ = 33 meV). For the 8${\times}$2 structure, ${\Delta}{\rho}$ shows a similar pattern with the 4${\times}$2 case (see Fig. 3 of the SI), yielding $E_{\rm g}$ = 0.19 eV and ${\Delta}E_{\rm 8{\times}2-4{\times}1}$ = $-$15 meV. We note that there is a general trend that the 8${\times}$2 structure is more stable than the 4${\times}$2 structure (see Table I). This indicates some energy gain caused by the correlation between two In nanowires in the 8${\times}$2 structure, as pointed out by a previous theoretical study~\cite{steko}.

\begin{figure}[ht]
\centering{ \includegraphics[width=7.0cm]{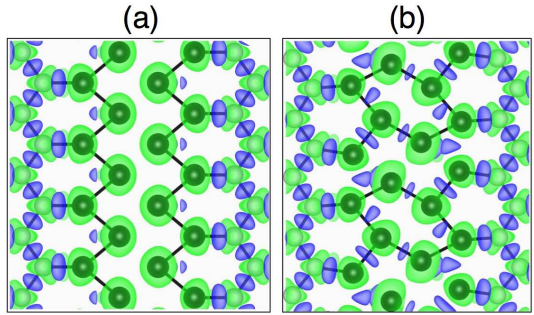} }
\caption{(Color on line) Charge density difference between ${\rho}_{\rm HSE}$ and ${\rho}_{\rm PBE}$ for the (a) 4${\times}$1
and (b) 4${\times}$2 structures. The dark (gray) color represents the isosurface of 0.02 ($-$0.02) e/{\AA}$^3$. }
\end{figure}

Using the HSE+vdW schme, we calculate the total energies of the 4${\times}$1, 4${\times}$2, and 8${\times}$2 structures with the PBE+vdW geometries~\cite{relax}. We find that the 4${\times}$2 and 8${\times}$2 structures are more stable than the 4${\times}$1 structure with ${\Delta}E_{\rm 4{\times}2-4{\times}1}$ = $-$23 meV and ${\Delta}E_{\rm 8{\times}2-4{\times}1}$ = $-$40 meV, respectively. Since the total energy is composed of the HSE energy ($E_{\rm HSE}$) and the vdW energy ($E_{\rm vdW}$) which is given by a sum of pairwise interatomic C$_6$$R^{-6}$ terms, the total energy difference between the 4${\times}$2 (or 8${\times}$2) and 4${\times}$1 structures is determined by
   \begin{equation}
    {\Delta}E = {\Delta}E_{\rm HSE}+{\Delta}E_{\rm vdW}.
   \end{equation}
Figure 4 shows ${\Delta}E_{\rm vdW}$ in ${\Delta}E_{\rm 4{\times}2-4{\times}1}$ and ${\Delta}E_{\rm 8{\times}2-4{\times}1}$, together with its components originating from In$-$In, In$-$Si, and Si$-$Si atoms.  We find that ${\Delta}E_{\rm HSE}$ in ${\Delta}E_{\rm 4{\times}2-4{\times}1}$ (${\Delta}E_{\rm 8{\times}2-4{\times}1}$) is $-$4 ($-$18) meV, while ${\Delta}E_{\rm vdW}$ in ${\Delta}E_{\rm 4{\times}2-4{\times}1}$ (${\Delta}E_{\rm 8{\times}2-4{\times}1}$) is $-$19 ($-$22) meV. Therefore, the inclusion of vdW interactions largely enhances the stabilization of the 4${\times}$2 and 8${\times}$2 structures. We note that the ${\Delta}E_{\rm HSE}$ values ($-$4 and $-$18 meV) in ${\Delta}E_{\rm 4{\times}2-4{\times}1}$ and ${\Delta}E_{\rm 8{\times}2-4{\times}1}$ are somewhat different from those (3 and $-$15 meV) obtained from the HSE calculation due to the use of two different PBE+vdW and PBE geometries. As shown in Fig. 4, the component of ${\Delta}E_{\rm vdW}$ originating from In$-$In atoms in the 4${\times}$2 and 8${\times}$2 structures is significantly larger in magnitude than those originating from In$-$Si and Si$-$Si atoms. Thus, we can say that the vdW interactions between In atoms play an important role in stabilizing the formation of hexagons.

\begin{figure}[ht]
\centering{ \includegraphics[width=7.0cm]{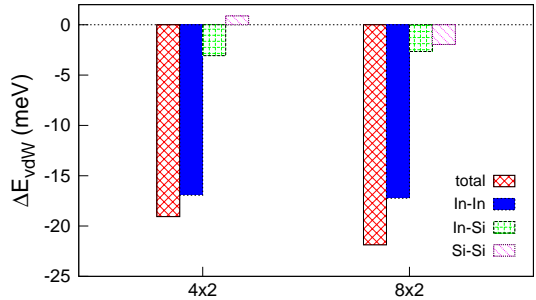} }
\caption{(Color on line) Contribution of vdW energy to the total-energy difference ${\Delta}E_{\rm 4{\times}2-4{\times}1}$ (${\Delta}E_{\rm 8{\times}2-4{\times}1}$) obtained using HSE+vdW. The components originating from In$-$In, In$-$Si, and Si$-$Si atoms are also given.}
\end{figure}

As shown in Fig. 2(d) [Fig. 1(d) in the SI], the HSE+vdW band structure of the 4${\times}$2 (8${\times}$2) structure gives $E_{\rm g}$ = 0.21 (0.21) eV, in good agreement with a recent scanning tunneling spectroscopy measurement of $E_{\rm g}$ = 0.2 eV~\cite{zhang}. We note that the HSE+vdW magnitude of ${\Delta}E_{\rm 8{\times}2-4{\times}1}$ is 40 meV per 4${\times}$1 unit cell, which is equal to 10 meV per In atom. Although the precise (4${\times}$1)${\leftrightarrow}$(8${\times}$2) phase transition temperature can be estimated by comparing the vibrational free energies of the 4${\times}$1 and 8${\times}$2 structures~\cite{wippermann}, the HSE+vdW magnitude of ${\Delta}E_{\rm 8{\times}2-4{\times}1}$ is well comparable with the thermal energy at the observed phase transition temperature of ${\sim}$120 K~\cite{yeom1,kump}. Therefore, the HSE+vdW scheme is likely to give good band gap and energetics of the low-temperature phase in the In/Si(111) system.

In summary, using the HSE and HSE+vdW schemes, we investigated the energy stability of the low-temperature and room-temperature structures in the In/Si(111) system, which has not been adequately described by previous DFT calculations~\cite{cho3,steko}. We found that the correction of SIE cures the delocalization error not only to give the insulating feature for the 4${\times}$2 and 8${\times}$2 structures but also to reverse the stability of the 4${\times}$1 and 8${\times}$2 structures. We also found that the vdW interactions between In atoms enhance the stability of hexagon structures. Our results demonstrate that the formation of hexagons in the In/Si(111) system occurs by a simple energy lowering due to the lattice distortion rather than by a Peierls instability. We notice that the Sn/Si(111) and Sn/Ge(111) surfaces have been the object of a large number of studies for determining the exact crystallographic arrangement, the electronic structure, and the mechanism of the phase transition~\cite{teje}. We anticipate that the correction of self-interaction error and the inclusion of vdW interactions would give more accurate description for the structural and electronic properties of such prototype two-dimensional electron systems.

This work was supported by National Research Foundation of Korea (NRF) grant funded by the Korean Government (NRF-2011-0015754). The calculations were performed by KISTI supercomputing center through the strategic support program (KSC-2012-C3-11) for the supercomputing application research.

\noindent $^{*}$ Corresponding author: chojh@hanyang.ac.kr



\begin{thebibliography}{99}

\bibitem{peierls} R. E. Peierls, "Quantum Theory of Solids", Clarendon: Oxford, (1964).
\bibitem{stewart} G. R. Stewart, Rev. Mod. Phys. {\bf 73}, 797 (2001).
\bibitem{jahn} H. A. Jahn and E. Teller, Proc. R. Soc London, Ser. A {\bf 161}, 220 (1937).
\bibitem{snijders} P. Snijders and H. Weitering, Phys. Mod. Phys. {\bf 82}, 307 (2010), and references therein.
\bibitem{bunk} O. Bunk, G. Falkenberg, J. H. Zeysing, L. Lottermoser, R. L. Johnson, M. Nielsen, F. Berg-Rasmussen, J. Baker, and R. Feidenhans'l, Phys. Rev. B {\bf 59}, 12228 (1999).
\bibitem{yeom1} H. W. Yeom, S. Takeda, E. Rotenberg, I. Matsuda, K. Horikoshi, J. Schaefer, C. M. Lee, S. D. Kevan, T. Ohta, T. Nagao, and S. Hasegawa, Phys. Rev. Lett. {\bf 82}, 4898 (1999).
\bibitem{kump} C. Kumpf, O. Bunk, J. H. Zeysing, Y. Su, M. Nielsen, R. L. Johnson, R. Feidenhans'l, and K. Bechgaard, Phys. Rev. Lett. {\bf 85}, 4916 (2000).
\bibitem{yeom2} H. W. Yeom, K. Horikoshi, H. M. Zhang, K. Ono, and R. I. G. Uhrberg, Phys. Rev. {\bf B 65}, 241307(R) (2002).
\bibitem{ahn} J. R. Ahn, J. H. Byun, H. Koh, E. Rotenberg, S. D. Kevan, and H. W. Yeom, Phys. Rev. Lett. {\bf 93}, 106401 (2004).
\bibitem{park} S. J. Park, H. W. Yeom, S. H. Min, D. H. Park, and I.-W. Lyo, Phys. Rev. Lett. {\bf 93}, 106402 (2004).
\bibitem{gon} C. Gonz\'alez, J. Guo, J. Ortega, F. Flores, and H. H. Weitering, Phys. Rev. Lett. {\bf 102}, 115501 (2009).
\bibitem{chandola} S. Chandola, K. Hinrichs, M. Gensch, N. Esser, S. Wippermann, W. G. Schmidt, F. Bechstedt, K. Fleischer, and J. F. McGilp, Phys. Rev. Lett. {\bf 102}, 226805 (2009).
\bibitem{cho1} J.-H. Cho, D. H. Oh, K. S. Kim, and L. Kleinman, Phys. Rev B {\bf 64}, 235302 (2001).
\bibitem{cho2} J.-H. Cho, J.-Y. Lee, and L. Kleinman, Phys. Rev B {\bf 71}, 081310(R) (2005).
\bibitem{tsay} S.-F. Tsay, Phys. Rev. B {\bf 71}, 035207 (2005).
\bibitem{lopez} X. L\'opez-Lozano, A. Krivosheeva, A. A. Stekolnikov, L. Meza-Montes, C. Noguez, J. Furthm\"uller, and F. Bechstedt, Phys. Rev. {\bf B 73}, 035430 (2006).
\bibitem{gon1} C. Gonz\'alez, J. Ortega, and F. Flores, New J. Phys. {\bf 7}, 100 (2005).
\bibitem{gon2} C. Gonz\'alez, F. Flores, and J. Ortega, Phys. Rev. Lett. {\bf 96}, 136101 (2006).
\bibitem{rii} S. Riikonen, A. Ayuela, and D. S\'anchez-Portal, Surf. Sci. {\bf 600}, 3821 (2006)
\bibitem{steko} A. A. Stekolnikov, K. Seino, F. Bechstedt, S. Wippermann, W. G. Schmidt, A. Calzolari, and M. Buongiorno Nardelli , Phys. Rev. Lett. {\bf 98}, 026105 (2007).
\bibitem{cho3} J.-H. Cho and J.-Y. Lee, Phys. Rev B {\bf 76}, 033405 (2007).
\bibitem{flores}
Gonz\'alez, J. Ortega, and F. Flores performed a first-principles tight-binding molecular dynamics calculation with a basis set of optimized atomic-like orbitals and the LDA using the FIREBALL code. Unlike the traditional use of the LDA which generally underestimates the band gap for insulators, their employed local orbital basis set with the LDA tends to overestimate band gaps [see Jel\'inek $et$ $al$., Phys. Rev. B {\bf 71}, 235101 (2005), and references therein], thereby leading to the stabilization of the insulating 4${\times}$2 and 8${\times}$2 structures over the metallic 4${\times}$1 structure.
\bibitem{ca} D. M. Ceperley and B. J. Alder, Phys. Rev. Lett. {\bf 45}, 566 (1980).
\bibitem{pw} J. P. Perdew, J. A. Chevary, S. H. Vosko, K. A. Jackson, M. R. Pederson, D. J. Singh, and C. Fiolhais, Phys. Rev. B {\bf 46}, 6671 (1992).
\bibitem{wippermann} S. Wippermann and W. G. Schmidt, Phys. Rev. Lett. {\bf 105}, 126102 (2010).
\bibitem{wall} S. Wall, B. Krenzer, S. Wippermann, S. Sanna, F. Klasing, A. Hanisch-Blicharski, M. Kammler, W. G. Schmidt, and M. Horn-vonHoegen, Phys. Rev. Lett. {\bf 109}, 186101 (2012).
\bibitem{hse} A. V. Krukau, O. A. Vydrov, A. F. Izmaylov, and G. E. Scuseria, J. Chem. Phys. {\bf 125}, 224106 (2006).
\bibitem{vdw} A. Tkatchenko and M. Scheffler, Phys. Rev. Lett. {\bf 102}, 073005 (2009).
\bibitem{aims} V. Blum, R. Gehrke, F. Hanke, P. Havu, V. Havu, X. Ren, K. Reuter, and M. Scheffler, Comput. Phys. Commun. {\bf 180}, 2175 (2009).
\bibitem{per} J. P. Perdew, K. Burke, and M. Ernzerhof, Phys. Rev. Lett. {\bf 77}, 3865 (1996); {\bf 78} 1396(E) (1997).
\bibitem{guo} G.-X. Zhang, A. Tkatchenko, J. Paier, H. Appel, and M. Scheffler, Phys. Rev. Lett. {\bf 107}, 245501 (2011).
\bibitem{sie1} J. P. Perdew and A. Zunger, Phys. Rev. B {\bf 23}, 5048 (1981).
\bibitem{sie2} S. K\"ummel and L. Kronik, Rev. Mod. Phys. {\bf 80}, 3 (2008).
\bibitem{relax} The PBE (PBE+vdW) geometry was kept fixed for the HSE (HSE+vdW) calculations. It was found that the fully relaxation of the geometries using HSE and HSE+vdW changes the energy difference between the 4${\times}$1 and 4${\times}$2 (or 8${\times}$2) structures by less than 3 meV per 4x1 unit cell.
\bibitem{zhang} H. Zhang, J.-H. Choi, Y. Xu, X. Wang, X. Zhai, B. Wang, C. Zeng, J.-H. Cho, Z. Zhang, and J. G. Hou, Phys. Rev. Lett. {\bf 106}, 026801 (2011).
\bibitem{teje} A. Tejeda, Y. Fagot-R\'evurat, R. Cort\'es, D. Malterre, E. G. Miche, and A. Mascaraque, Phys. Status Solidi A {\bf 209}, 614 (2012), and references therein.
\end{thebibliography}
\end{document}